\documentclass[10pt, letterpaper]{article}
\usepackage{opex3}
\usepackage{cite}
\usepackage{color, float}

\begin{document}

\title{Noncollinear parametric fluorescence by chirped quasi-phase matching for monocycle temporal entanglement}

\author{Akira Tanaka,$^{1,2}$ Ryo Okamoto,$^{1,2}$ Hwan Hong Lim,$^{3}$ Shanthi Subashchandran,$^{1,2}$ Masayuki Okano,$^{1,2}$ Labao Zhang,$^{4}$ Lin Kang,$^{4}$ Jian Chen,$^{4}$ Peiheng Wu,$^{4}$ Toru Hirohata,$^{5}$ Sunao Kurimura,$^{3}$ and Shigeki Takeuchi$^{1,2,*}$}

\address{${}^1$ Research Institute for Electronic Science, Hokkaido University, Kita-ku, Sapporo 001-0020, Japan\\
${}^2$ The Institute of Scientific and Industrial Research, Osaka University, 8-1 Mihogaoka, Ibaraki, Osaka 567-0047, Japan\\
${}^3$ National Institute for Materials Science, 1-1 Namiki, Tsukuba, 305-0044, Japan\\
${}^4$ Research Institute of Superconductor Electronics (RISE), School of Electronic Science and Engineering, Nanjing University, Nanjing 210093, China\\
${}^5$ Central Research Laboratory, Hamamatsu Photonics, K.K., 5000 Hirakuchi, Hamamatsu City, 434-8601, Japan}

\email{*takeuchi@es.hokudai.ac.jp} 

\begin{abstract}
Quantum entanglement of two photons created by spontaneous parametric downconversion (SPDC) can be used to probe quantum optical phenomena during a single cycle of light.  Harris [Phys. Rev. Lett. \textbf{98}, 063602 (2007)] suggested using ultrabroad parametric fluorescence generated from a quasi-phase-matched (QPM) device whose poling period is chirped.  In the Harris's original proposal, it is assumed that the photons are collinearly generated and then spatially separated by frequency filtering.  Here, we alternatively propose using noncollinearly generated SPDC. In our numerical calculation, to achieve 1.2 cycle temporal correlation for a 532 nm pump laser, only 10\% -chirped device is sufficient when noncollinear condition is applied, while a largely chirped (50\%) device is required in collinear condition.  We also experimentally demonstrate an octave-spanning (790-1610 nm)  noncollinear parametric fluorescence from a 10 \% chirped MgSLT crystal using both a superconducting nanowire single-photon detector and photomultiplier tube as photon detectors.  The observed SPDC bandwidth is 194 THz, which is the largest width achieved to date for a chirped QPM device. From this experimental result, our numerical analysis predicts that the bi-photon can be compressed to 1.2 cycles with appropriate phase compensation.
\end{abstract}



\section{Introduction}
Quantum entanglement plays a key role in various systems, such as quantum computation \cite{2} and metrology \cite{3}.  An example is a pair of photons generated by parametric downconversion, which is a second-order nonlinear optical process\cite{4,5}.  The temporal entanglement of the two photons can show a simultaneity in their arrival times to within their time-bandwidth product, which cannot occur for classical light \cite{21}.  If two photons arrive within several femtoseconds of each other, with one cycle equal to the inverse of its center frequency (1 cycle=$1/\nu_c$), potential applications include an enhancement in the resolution of quantum optical coherence tomography \cite{6,7} and in the rate of two-photon absorption or sum-frequency generation \cite{8,9,Kurimura2}. 

To realize ultrashort temporally entangled two-photon states,  the parametric fluorescence must have a bandwidth $\Delta\nu$ that is comparable to its center frequency $\nu_c=c/\lambda_c$, e.g., 282 THz at a center wavelength of $\lambda_c=1064$ nm.  In bulk nonlinear crystals, broadband phase matching within a small solid angle of emission can only occur over a short interaction length, thereby degrading the photon pair generation rate.  To date, such pair emission has been limited to a bandwidth of $\Delta\nu =154$ THz (253 nm) at $\lambda_c$ = 702 nm by introducing a temperature gradient in a crystal \cite{kulik2}.  We have recently achieved $\Delta\nu$=73 THz (160 nm) at $\lambda_c$ = 808 nm using two bulk crystals with their optic axes tilted relative to each other \cite{okano}. 

Harris \cite{1} proposed a method to compress the temporal correlation of biphotons to a single optical cycle of light (e.g. 3.5 fs at a wavelength of 1064 nm).  The biphotons propagate collinearly with a large frequency bandwidth due to an aperiodically poled nonlinear optical crystal, i.e. a chirped quasi-phase-matched (QPM) device\cite{Fejer}.  A photon pair which is phase matched at each poling period possesses different combinations of two frequencies, resulting in an ultra-broadband two-photon state after passing through the crystal.  After spatially splitting the photons and applying a phase compensation to one path and a time delay to the other, the temporal width of the biphotons was  measured by sum-frequency generation (SFG).  Nasr et al. has demonstrated a bandwidth increase using such a device at  $\lambda_c$ = 808 nm to achieve $\Delta\nu$=188 THz for a two-photon interference experiment\cite{12}.  Mohan et al. has shown spectral broadening at  $\lambda_c$ = 1064 nm, resulting in $\Delta\nu$=166 THz in a collinear condition\cite{7}.  Sensarn et al. has succeeded in a "chirp and compress" experiment of  at $\lambda_c$ = 1064 nm and $\Delta\nu$=40 THz at a nondegenerate lobe \cite{11}.  

However, there are some difficulties.  In Harris's original proposal, which uses a collinear condition and require very large bandwidth ($\Delta\nu$=3000 nm) i.e., a QPM device with large chirping (50 \%) for 1.2 cycle temporal correlation when a pump laser at 532 nm is used (see section 2 for detail).  The realization of such a device is still beyond current technologies.  In addition, the dichroic mirror used to separate the fluorescence to lower and higher frequencies also causes additional problems like anomalous dispersion near the cutoff wavelength. 

In the present paper, first we propose using noncollinear SPDC. We show theoretically that the chirp rate to achieve monocycle temporal correlation can be substantially reduced compared with collinear SPDC\cite{1}. Next, we experimentally demonstrate an octave-spanning (790-1610 nm) noncollinear parametric fluorescence from a 10 \% chirped MgSLT crystal using both a superconducting nanowire single-photon detector and a photomultiplier tube. Finally, we report our numerical calculation, which suggests that the temporal correlation of the two photon can be 1.2 cycle with a perfect chirp compensation, and to 7.25 cycle with a moderate chirp compensation using a prism pair, assuming the experimentally observed spectra.

To measure two-photon temporal correlations using an SFG crystal, the full range of frequencies is integrated in contrast to collinear scheme, yielding the absolute square of the two-photon wavefunction in the time domain (for ideal phase matching).  As is shown later, the same ultrashort time correlation of 1.2 cycle is attainable with a QPM device with just 10 \% chirp.  This result relaxes the requirement for the chirp rate of a chirped-QPM device to reach the monocycle regime.  

Towards the realization of the scheme using noncollinear condition, we experimentally implemented direct spectroscopy of ultra-broadband noncollinear parametric fluorescence from a chirped-QPM device using a wavelength filter and two detectors: a superconducting nanowire single-photon detector (SNSPD)\cite{Shanthi} and a photomultiplier tube (PMT)\cite{Hirohata}.  These detectors span the bandwidth of generated ultra-broad single-photon wavepackets across an octave.  A chirped-QPM device is fabricated whose poling period is chirped from 8.000 to 8.825 $\mu$m along the pump direction, corresponding to 10 \% chirping.  The spectrum thus obtained for both the signal and idler beam has a bandwidth of 194 THz, the largest so far observed for visible light pumping (at 532 nm).  

To check the feasibility of the future realization of monocycle entangled states, the effects of phase compensation on the spectral phase of two-photon states is calculated.  We assumed two cases; using a prism pair and a full dispersion compensation which may be possible using a spatial light modulator\cite{11,10}.  For perfect compression, the device can attain 1.2 cycles.  Even when a prism pair is used as a dispersive element, compression by a factor of 7.54$\times 10^{-3}$ can be achieved and 7.25 cycle signal may be obtained.  In other words, the parametric fluorescence with appropriate phase compensation can produce monocycle entangled photon pairs.  These results will open the way for ultrafast two-photon processing technologies\cite{12} such as submicron quantum optical coherence tomography\cite{6,7} and nonclassical light generation via electromagnetically induced transparency\cite{du}.  

The rest of the paper is organized as follows.  The structural aspects of a chirped quasi-phase-matched device and the theory of noncollinear two photon states are described in Sec 2.  An experimental setup using two complementary broadband detectors is demonstrated in Sec 3.  In Sec 4, results obtained using both an SNSPD and a PMT spanning more than an octave in a noncollinear condition are presented.  In Sec 5, numerical results for the temporal correlation seen in ideal sum-frequency generation measurements is compared to realistic measurements using prism pairs.

\section{Generation and measurement of ultrabroad two-photon states}
\subsection{Chirped quasi-phase-matched device for generation of two entangled photons}
The chirped quasi-phase-matched device is shown in Fig. 1.  The crystal consists of slabs with different poling periods to achieve the chirp\cite{1,Fejer2}
\begin{equation}
K(z)\equiv \frac{2\pi}{\Lambda (z)}=\frac{2\pi}{\Lambda_0}-\eta z
\end{equation}
\begin{figure}[b]
\centering\includegraphics[width=13cm]{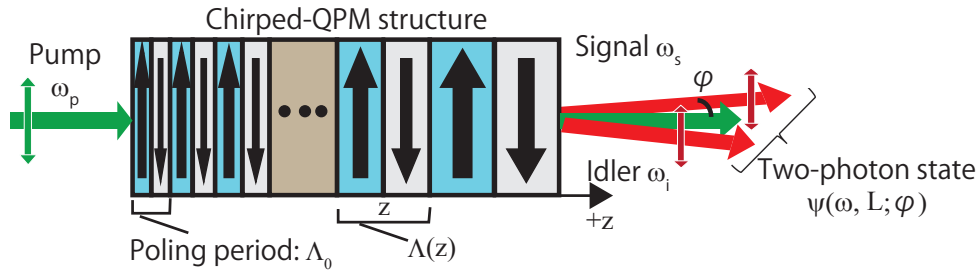}
\caption{(a) Schematic of the chirped quasi-phase-matched device. }
\end{figure}
where the $z$ axis is along the pump beam, with origin at the input face of the crystal; $K(z)$ is the spatial wavevector with poling period $\Lambda (z)$, where $\Lambda_0$ is the poling period at the input face; and $\eta$ is the chirp rate increasing the poling period as $z$ increases.  The noncollinear SPDC process converts the pump photons at an angular frequency $\omega_p$ into signal ($\omega_s$) and idler ($\omega_i$) photons propagating at angle $\varphi$ relative to the pump.  Conservation of energy and momentum\cite{Baek} is expressed as
\begin{equation}
\omega_p=\omega_s+\omega_i
\end{equation}
\begin{eqnarray}
\Delta k(\omega, z;\varphi)=k(\omega_p)-k(\omega)\sqrt{1-\left(\frac{\sin{\varphi}}{n_e(\omega)}\right)^2}\nonumber \\
-k(\omega_p-\omega)\sqrt{1-\left(\frac{\omega}{\omega_p-\omega}\right)^2\left(\frac{\sin{\varphi}}{n_e(\omega_p-\omega)}\right)^2}-K(z) =0
\end{eqnarray}
where $k(\omega)=n_e(\omega)\omega /c_0$ is the wavevector component with extraordinary refractive index $n_e(\omega)$ and speed of light $c_0$, and $\Delta k(\omega,z;\varphi)$ is the phase mismatch.  The coupled wave equation for the two-photon wavefunction at the output face of the crystal is\cite{1,wavefunction}
\begin{eqnarray}
 \psi (\omega,L;\varphi) = -\sqrt{\frac{i \kappa^2 \pi}{2\eta}}e^{i\{ k(\omega)+k(\omega_p-\omega) \} L}e^{-i\Delta k(\omega,0;\varphi)^2/2\eta}
\nonumber \\
\times\left[ f\left( \frac{1+i}{2\sqrt{\eta}}\Delta k(\omega,0;\varphi)\right)-f\left( \frac{1+i}{2\sqrt{\eta}}(\Delta k(\omega,0;\varphi)+\eta L)\right)\right]
\end{eqnarray}
where $\kappa$ is the parametric fluorescence generation efficiency and $f(x) =  \frac{-2i}{\sqrt{\pi} }\int_0^{ix} e^{-t^2}dt$ is the imaginary error function.  The average photon number per mode is $|\psi (\omega,L;\varphi)|^2/{2\pi}| $\cite{1}, used to compare experimental results to theory for the collinear($\varphi=0$) and noncollinear cases.

\subsection{Comparison of entangled photons in the collinear and noncollinear cases}
The quantum states for collinear and noncollinear SPDC from a chirped-QPM device, denoted $|\Phi\rangle_{col}$ and $|\Psi\rangle_{noncol}$, are
\begin{eqnarray}
|\Phi\rangle_{col} =\int_{-\infty}^{\infty}d\omega  \psi (\omega,L;\varphi=0) |\omega,\omega_p-\omega\rangle_{col} \\
|\Psi\rangle_{noncol}=\int_{-\infty}^{\infty}d\omega  \psi (\omega,L; \varphi) |\omega\rangle_s|\omega_p-\omega\rangle_i 
\end{eqnarray}
where $|\rangle_{col}$ corresponds to a spatial mode of the biphoton and $|\rangle_{s(i)}$ that of the signal (idler) state of the entangled two-photon state.

\begin{figure}[t]
\centering\includegraphics[width=13cm]{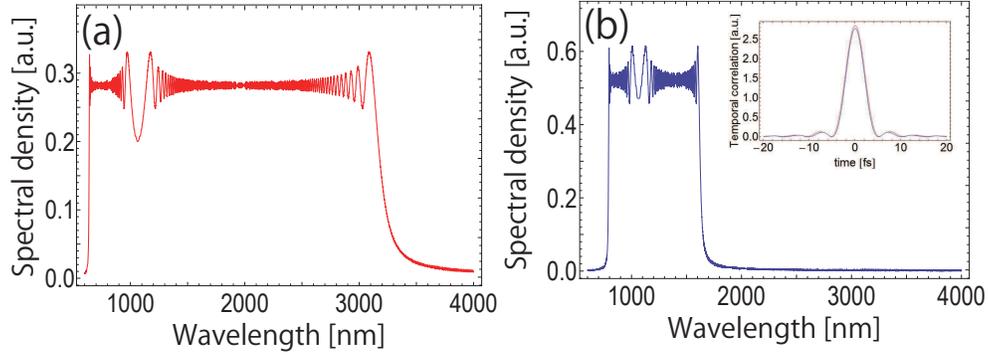}
\caption{Simulations of  (a) collinear (47 \% chirp)  and (b) noncollinear(10 \% chirp) parametric fluorescence  spectra to achieve monocycle temporal correlation.  Inset: calculation of the expected SFG signals of  $R_{SFG}^{col}(\tau)$ and  $R_{SFG}^{noncol}(\tau)$ for a two-photon state with a FWHM of 4.4 fs based on spectra in (a) and (b).}
\end{figure}
In the scheme by Harris \cite{1}, they split biphotons into two groups by passing photons at $\omega\le\omega_p/2$ while reflecting those at $\omega \ge\omega_p/2$ using a frequency filter, one of which passes through a dispersive element $H(\omega)$ and the other experiences a time delay $G(\omega_p-\omega; \tau)$.  The two are then combined in an SFG crystal.  The resulting signal for ideal phase matching is 
\begin{equation}
R_{SFG}^{col}(\tau)\propto\left|\frac{1}{2\pi}\int_{\omega_p/2}^{\infty}H(\omega )G(\omega_p-\omega; \tau ) \psi (\omega,L;\varphi=0)e^{i\omega\tau}d\omega\right|^2
\end{equation}  
in the collinear case.  For the noncollinear case, since each photons need not to be separated by such wavelength filters in order to experience either $H(\omega)$ or $G(\omega_p-\omega; \tau)$,
\begin{equation}
R_{SFG}^{noncol}(\tau)\propto\left|\frac{1}{2\pi}\int_{-\infty}^{\infty}H(\omega )G(\omega_p-\omega; \tau ) \psi (\omega,L;\varphi)e^{i\omega\tau}d\omega\right|^2.
\end{equation}  
For $R_{SFG}^{col}(\tau)$, the lower limit of integration is $\omega_p/2$ to divide biphotons in two pathways by a dichroic mirror.   If the complex phase of $\psi (\omega,L;\varphi)$ could be compensated with $H(\omega)$, i.e. $\psi (\omega,L;\varphi)H(\omega)$ being real in the loss-free scenario, then $R_{SFG}^{noncol}(\tau)\propto|\tilde{\Psi}(\tau;\varphi)|^2$ where $\tilde{\Psi}(\tau;\varphi)$ is the two-photon wavefunction in the time domain, according to the Fourier transform in Eq. (8).  
The frequency bandwidth to reach an SFG temporal width of 4.4 fs is calculated from Fig. 2, using Eqs. (7) and (8) and the Sellmeier equation for MgO-doped stoichiometric lithium tantalate at 293 K\cite{Sellmeier}.  In Fig. 2(a), the fluorescence spectrum spans from 650 to 3500 nm with a poling period that increases from $\Lambda_0=8.000\mu m$ to $11.765\mu m$, i.e., 47 \% chirping.  In contrast, the noncollinear fluorescence spectrum for $\varphi=0.25$deg in Fig. 2(b) only spans from 790 to 1610 nm with a poling period that increases from $\Lambda_0=8.000\mu m$ to $8.825\mu m$, i.e., a chirp rate of 10 \%, relaxing the difficulty of creating the structure. The inset of Fig. 2(b), shows the SFG signal for the collinear or noncollinear cases, which are almost identical. Also, when we set equal frequency bandwidths for SFG as in Fig. 2(b), the FWHM is only 8.8 fs (2.5 cycle) for collinear SPDC while noncollinear SPDC gave 4.4 fs (1.2 cycle). 

Let us explain why the correlation time using noncollinear SPDC is smaller than that using collinear SPDC. The correlation time is given roughly by the inverse of the bandwidth of the daughter photons in two separate optical paths. For collinear SPDC, one has to separate two daughter photons generated in the same optical mode into two optical paths. In HarrisE's proposal\cite{1}, a dichroic filter was used to separate the photons, however, the bandwidth of each of the photons become became a half of the original one. On the other hand, for noncollinear SPDC, the two photons are generated in two different optical paths from the beginning and thus one does not have to use such a dichroic filter. Therefore, the bandwidth of these photons in two paths is twice larger than that using collinear SPDC. This is the reason why the correlation time using noncollinear SPDC is smaller than that using collinear SPDC.

\section{Experimental setup and fabrication of chirped-QPM device}
\subsection{Fabrication of the device}
The chirped-QPM device is based on 1.0 mol\% magnesium-doped stoichiometric lithium tantalate (MgSLT) with a Curie temperature of 685 celsius Celsius\cite{Kurimura}.  The crystal was grown by the double-crucible Czochralski method for high compositional uniformity.  Its ferroelectruc ferroelectric spontaneous polarization is aperiodically switched by electric field poling with patterned electrodes.  Its dimensions are 0.5 mm $\times$ 0.5 mm in cross section for good heat removal and 20 mm in length for significant photon count in SPDC.  The small cross sections enables us to obtain stable SPDC spectrum due to the suppressed thermal effect even at high pump power\cite{Lim}.  The patterned periodic structure has a chirp rate of $\eta$=367.112 rad$\cdot$cm${}^{-2}$ corresponding to $\Lambda_0$=8.000 $\mu$m.  The device is designed for a first-order QPM in the type-zero SPDC condition in which all photons consist of extraordinary rays, as shown in Fig. 1.  This device, designed for a cw pump at a wavelength of 532 nm, is temperature controlled by a Peltier unit at 293 K. 
\begin{figure}[b]
\centering\includegraphics[width=11cm]{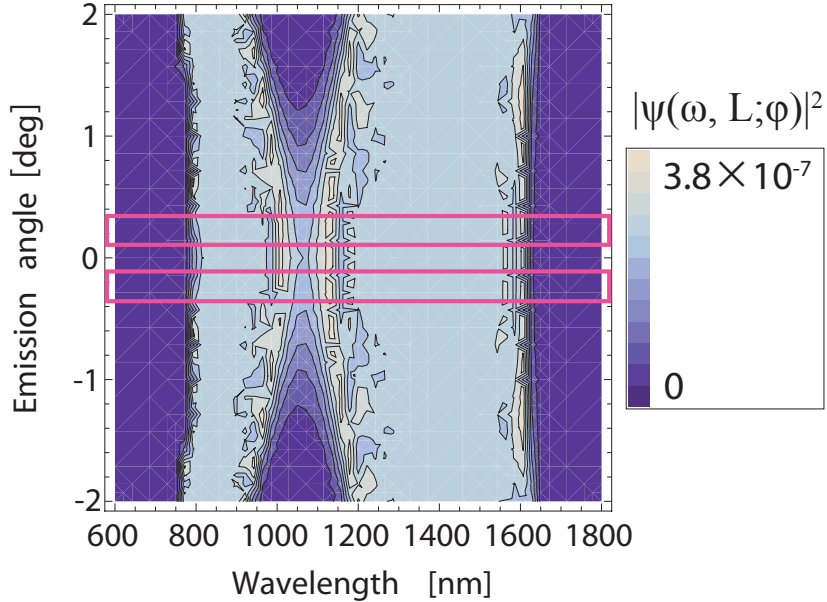}
\caption{Calculated tuning curve for Type-0 SPDC from the chirped-QPM device. The parametric fluorescence in the experiment lies within the two pink boxes centered at angles of $\pm$0.25 deg and spanning a width of 0.28 deg. }
\end{figure}

\subsection{Calculation of the tuning curve}
An angle-resolved spectrum of  the parametric fluorescence, serving as tuning curve, is shown in Fig. 3 based on the analytical expression of Baek and Kim \cite{Baek}.  The term $|\Delta k(\omega, L; \varphi)|^2$ in Eq. (4) is plotted as a function of wavelength $\lambda$ and emission angle $\varphi$.  The horizontal axis of the tuning curve gives the wavelength of the photons and the vertical axis the emission angle of the photons into air.  Note that in the contour map, the value at specific point expresses the square of the two-photon wavefunction for the combination of $\lambda$ and $\theta$\cite{wavefunction}.
The noncollinear angles are $\varphi=0.25\pm 0.14$ degree and $-0.25\pm0.14$ degree, each marked by a red box in Fig. 3, for which no significant change occurs in the spectrum compared to the collinear case.  In the following experiment, we used an iris having two horizontal apertures (2 mm in diameter) to limit the spatial mode.  The alignment of the iris was accomplished by observing the parametric fluorescence from QPM device at 304 K with a CCD camera (Princeton Instruments PIXIS 1024 BTSM).

\subsection{Experimental setup}
\begin{figure}[t]
\centering\includegraphics[width=13cm]{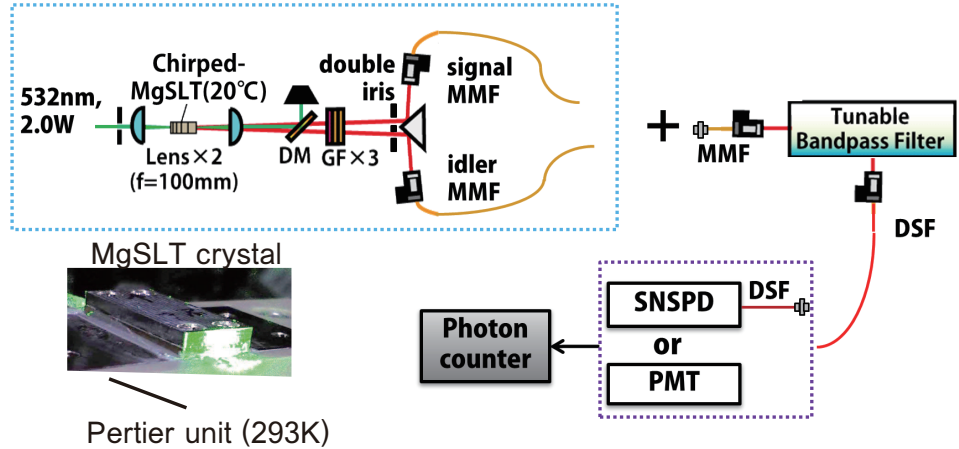}
\caption{ Experimental setup for the noncollinear geometry.  MMF: multi-mode fiber, DSF: dispersion-shifted fiber, SNSPD: superconducting single photon detector, PMT: photomultiplier tube.  Inset: photograph of the chirped-QPM device on a peltier unit.}
\end{figure}
Figure 4 illustrates the setup for the detection of parametric fluorescence emitted from the chirped-QPM device in a noncollinear condition.  A diode-pumped solid-state laser (Coherent Verdi-V10 at 532 nm and 10 W) is used as a pump to generate the fluorescence.  The pump power  is adjusted using polarization components and set to 2W in front of the crystal.  Two lenses (having a focal length of 100 mm) are used to focus the pump light onto the MgSLT crystal and collimate the fluorescence.  The pump beam was attenuated to 75\% (1.5 watt) of the input after the MgSLT crystal. 

The pump light is reflected by a dichroic mirror having an optical density of 6 (Semrock LP02-568RS-25) while the fluorescence is transmitted.  Following three glass filters, to further reduce the residual pump, a prism mirror splits the photons into two paths and couples them into different multimode fibers.  A bandpass filter (Photon etc. Laser Line Tunable Filter) tunes the wavelength of the output light between 600 and 1800 nm.  The FWHM of the output light is 4/6 nm below/above 1.1 $\mu$m, due to the use of two Bragg gratings.  The output light is coupled into a dispersion-shifted fiber having a mode field diameter of 8.0 $\mu$m in wavelength of 1550 nm.  The photons are then routed either to a superconducting nanowire single photon detector (SNSPD) \cite{Shanthi} or to a photomultiplier tube (PMT). Both detectors create a pulse that is measured by a photon counter (Stanford Research Systems SR400) with an integration time of 1 s.

\section{Spectrum of the noncollinear parametric fluorescence}
\subsection{Experiment with SNSPD}
We have developed a meander Superconducting Nanowire Single Photon Detector (SNSPD) consisting of a NbN nanowire at 4.2 K \cite{Shanthi}.  It has a dark count of only 1.7$\pm$ 1.3 counts/s at a critical bias current of 37.2 $\mu$A.  Single-photon detection from at least 500 to 1650 nm is possible, however, the quantum efficiency exponentially decreases as a function of wavelength (30.7\% at 600 nm, 16.6\% at 800 nm, 10.3\% at 1000 nm, 3.3\% at 1300nm and 1.1\% at 1550 nm with a bias current of 0.95 Ic where Ic is a critical current).  For this reason, it is necessary to test its performance in the NIR wavelength range. 

\begin{figure}[t]
\centering\includegraphics[width=13cm]{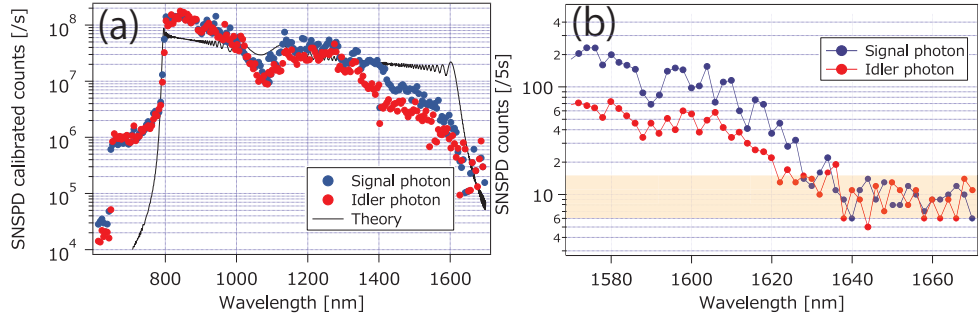}
\caption{(a) Measured parametric fluorescence spectra after calibration for noncollinear propagation detected by an SNSPD, together with a theoretical curve including the wavelength resolution of the tunable bandpass filter. We inserted a long wavelength pass filter for the wavelength longer than 1400 nm. (b) The raw data of the fluorescence spectra at longer wavelength region (accumulation time: 5sec) after inserting the same long wave pass filter.  Pale shadow: typical dark counts.}
\end{figure}
\begin{figure}[h]
\centering\includegraphics[width=8cm]{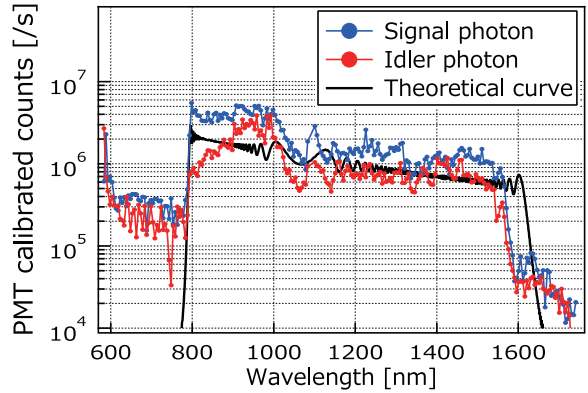}
\caption{Logarithmic plots of the parametric fluorescence spectra for noncollinear propagation detected by the PMT after intensity calibration. The black line is a theoretical curve accounting for the wavelength resolution of the tunable bandpass filter.}
\end{figure}

In Fig. 5(a), the parametric fluorescence spectrum emitted in the noncollinear condition is graphed after calibrating the detection efficiency of SNSPD, the coupling efficiency of pump light into dispersion shifted fiber through bandpass filter and transmittance spectrum of bandpass filter. The blue and red curves plot the photon counts emitted at $\pm$0.25 degrees.
Note that due to the (unwanted) second order diffraction of the tunable band-pass filter, photons at the wavelength around 800nm may become non-negligible noise in the spectrum around 1600 nm since the detection efficiency of SNSPD at 800 nm is much larger than that at 1600 nm. Thus, we inserted a long-wavelength pass filter (Newport 10LWF-1000-B) for the wavelength longer than 1400nm to remove this noise. For this region, the transmittance (typ. 87 \%) of the long-wavelength pass filter was calibrated.
The black line is a theoretical curve integrated over the acceptance angle and multiplied by the resolution of the bandpass filter ($\delta\lambda$: set approximately to 5 nm), i. e. $\int_{0.11deg}^{0.39deg} d\varphi |\Psi (\omega, L;\varphi)|^2 d\omega \delta\omega$ with $\Psi (\omega, L;\varphi)$ in Eq. (4) where $\delta\omega$ is the effective resolution in the angular frequency region..  It has then been scaled vertically to match the data.  The theoretical curve increases at shorter wavelength due to fixed wavelength resolution $\delta\lambda$ that changes $\delta\omega$ in theoretical curve.  The wavelength range from 790 to 1610 nm is in reasonable agreement with the numerical prediction. Moreover, we have confirmed in Fig. 5(b) that the longer wavelength range actually spans until 1610 nm with the same experimental condition except the accumulation time of 5 sec. On the other hand, one can see a discrepancy between the estimated photon counts at shorter wavelengths and theoretical curve in Fig. 5(a) from 800nm to 1000nm.  This difference may originate from the wavelength dependence of the coupling efficiency into the multi-mode fiber due to the aberration in the objective lens.  The photon counts drop from approximately $2\times 10^4$ counts/s at 800 nm to only 140 counts/s at 1600 nm.  This large change arises from the exponential decrease in the quantum efficiency.  Below 790 nm, additional noise arises due to stray light from the pump laser which is smaller than SPDC by two orders of magnitude.  Nevertheless, the generated photon pairs exhibit spectral indistinguishability in their amplitudes; only the phase has to be compensated to compress biphotons into a single cycle, used to achieve two-photon temporal entanglement \cite{18}.

\subsection{Experiment with PMT}
\begin{figure}[b]
\centering\includegraphics[width=10cm]{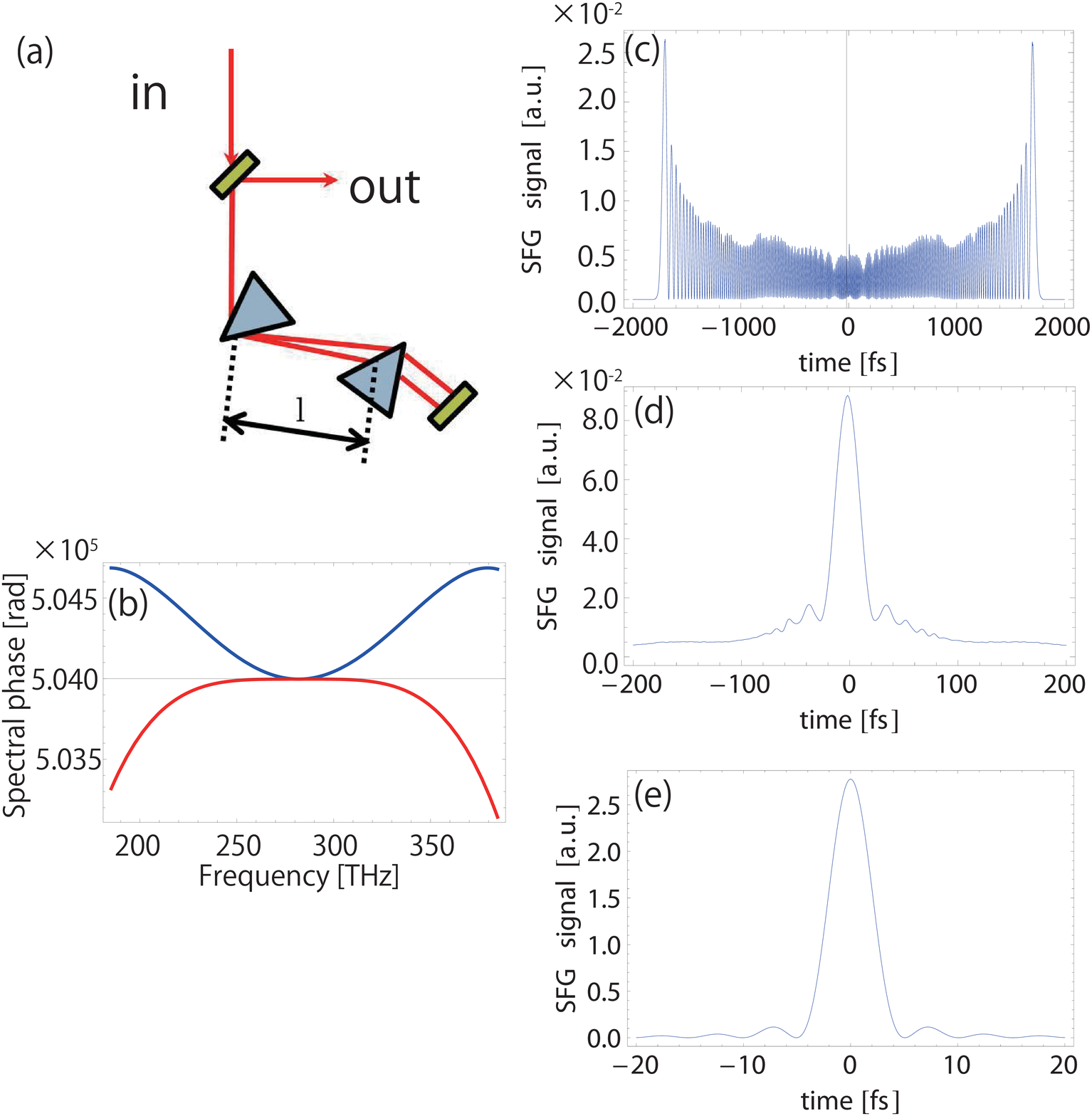}
\caption{(a) Schematic of the prism compressor with a separation of $l$. (b) Simulated spectral phase of the two-photon wavefunction before (blue) and after (red) insertion of the prism compressor. SFG signals (c) before and (d) after insertion of the compressor, having a FWHM of 3600 fs (1000 cycles) and 26.6 fs (7.5 cycles).  (e) SFG signal after perfect compression with a FWHM of 4.4 fs, corresponding to 1.1 1.2 cycles. }
\end{figure}
To check the spectral reliability, an InP/GaAs/InP photomultiplier tube (PMT) can be used because it has a flat quantum efficiency from 350 to 1550 nm compared with SNSPD, although the detector has a large dark noise of 2.2$\times 10^4$ counts/s.    Figure 6 shows a calibrated logarithmic plot of parametric fluorescence measured by a PMT after removal of all glass filters. They are in good agreement with the numerical results, except near 1600 nm since the quantum efficiency of the PMT suddenly decreases from 1550 nm.  As is evident from these results, the flatness of quantum efficiency suppressed the fluctuation of the calibrated counts at longer wavelength range, compared with SNSPD data in which small photon counts are multiplied in calibration procedure. Note that unlike Fig. 5, we did not use a long-wavelength pass filer for Fig. 6 because the second-order diffracted photons at the wavelength around 800nm are negligible due to the flat quantum efficiency of PMT for a wide range of wavelength.

\section{Effects of spectral phase compensation on the two-photon state}

Since photon pairs generated at different locations in a chirped-QPM crystal experience different dispersion, and therefore generate spectrally distinguishable information, the temporal correlation between the signal and idler photons is not directly related to the inverse bandwidth.  Photon pairs are emitted with the spectral phase of Eq. (3) plotted in blue in Fig. 7(b),
\begin{equation}
\phi_{spec}(\omega)\equiv k(\omega)L+k(\omega_p-\omega)L-[\Delta k(\omega,0)]^2/2\eta.
\end{equation}
As indicated by the SFG signals (related to the two-photon wavefunction in the time domain $|\tilde{\psi}(\tau)|^2$) in Fig. 7(c), without phase compensation the temporal correlation of the photons is extended to 3.6 ps from the transform-limited width of 4.4 fs, corresponding to 1000 cycles, an increase in three orders of magnitude.  The horizontal axis is the arrival time difference $\tau$ between signal and idler photons at an SFG crystal and the vertical axis is $|\tilde{\Psi}(\tau)|^2$, proportional to $R^{noncol}(\tau)$ in Eq. (8).  The spectral phase can be partially eliminated by a standard pulse compression technique on the signal or idler photons.  The prism compressor in Fig. 7(a) \cite{Trebino} removes unwanted group delay dispersion $d^2\phi_{spec}(\omega)/d\omega^2$=7.4$\times$10${}^3$ fs${}^2$ intrinsic to our device, by using two SF14 prisms with a separation of 500 mm.  The red curve in Fig. 7(b) plots the simulated spectral phase of the biphotons following the prism compressor.  As a result, a reduction in phase difference is expected around the degenerate frequency of 282 THz.  However, due to higher even-order contributions of the spectral phase, the SFG signal cannot reach the monocycle regime, yielding 26.6 fs (7.5 cycle) in Fig. 7(d) \cite{19}.  To reduce such residual phase and additional odd-order components from the prism compressor, a spatial phase modulator in a 4$f$-optical system should work, reaching a Fourier transform-limited width of 4.4 fs in Fig. 7 (e) \cite{10}.

\section{Conclusion}
We showed theoretically that using noncollinear SPDC, we can reduce the chirp rate necessary to achieve monocycle temporal correlation compared with collinear SPDC\cite{1}. We also experimentally realized an octave-spanning (790-1610 nm) noncollinear parametric fluorescence from a 10 \% chirped MgSLT crystal, which was observed using both a superconducting nanowire single-photon detector and a photomultiplier tube. Finally, we reported our numerical calculation, which suggests that the temporal correlation of the two photon can be 1.2 cycle with a perfect chirp compensation, and to 7.25 cycle with a moderate chirp compensation using a prism pair, assuming the experimentally observed spectra.

\section*{Acknowledgments}
We thank Prof. Mikio Yamashita for kind discussion and Mr. Takahiro Shimizu for teaching device fabrication.  This work is supported by JST-CREST, Quantum Cybernetics, JSPS-FIRST, the Japanese Society for the Promotion of Science, the Research Foundation for Opto-Science and Technology, Special Coordination Funds for Promoting Science and Technology, a Grant-in-Aid for JSPS Fellows (11J00744), JSPS Research Fellowships for Young Scientists, and the G-COE program.


\begin{thebibliography}{99}
\bibitem{19}  Note that the spectral phase of biphotons is mostly an even function of $\omega_p/2$; because a Hong-Ou-Mandel interferometer cancels even order dispersion, the HOM dip that appears in reference \cite{12} using a chirped PPSLT crystal leads to a FWHM of 4 fs, which is nearly Fourier-transform limited.
\end{thebibliography}
\end{document}